\documentclass{article}
\usepackage{amsmath}

\usepackage{arxiv}

\usepackage[utf8]{inputenc} 
\usepackage[T1]{fontenc}    
\usepackage{hyperref}       
\usepackage{url}            
\usepackage{booktabs}       
\usepackage{amsfonts}       
\usepackage{nicefrac}       
\usepackage{microtype}      
\usepackage{lipsum}
\usepackage{graphicx}
\graphicspath{ {./images/} }
\usepackage{cleveref}
\crefname{figure}{Figure}{Figures}
\Crefname{figure}{Figure}{Figures}
\usepackage{multirow}
\usepackage{array}
\usepackage{float}
\usepackage{multirow, booktabs}
\usepackage{makecell}
\usepackage{url}

\newcommand{\hnmr}{\textsuperscript{1}H} 
\newcommand{\cnmr}{\textsuperscript{13}C}

\usepackage{multicol} 
\usepackage{authblk}

\title{DiffNMR: Diffusion Models for Nuclear Magnetic Resonance Spectra Elucidation}

\author[1]{Qingsong Yang\textsuperscript{*}}   
\author[2]{Binglan Wu\textsuperscript{*}}
\author[3]{Xuwei Liu}
\author[2]{Bo Chen}
\author[2]{Wei Li}
\author[2]{Gen Long\textsuperscript{†}} 
\author[2]{Xin Chen\textsuperscript{†}} 
\author[1]{Mingjun Xiao\textsuperscript{†}}     

\affil[1]{Department of Computer Science, University of Science and Technology of China}
\affil[2]{Suzhou Lab}
\affil[3]{Baidu}

\let\oldmaketitle\maketitle
\renewcommand{\maketitle}{%
  \oldmaketitle
  \vspace*{-1em} 
  \footnotetext[1]{* Equal contribution} 
  \footnotetext[2]{\textsuperscript{†} Corresponding authors: mingjun.xiao@ustc.edu.cn, xin.chen@suzhou.ac.cn}
}

\begin{document}
\maketitle
\begin{abstract}
Nuclear Magnetic Resonance (NMR) spectroscopy is a central characterization method for molecular structure elucidation, yet interpreting NMR spectra to deduce molecular structures remains challenging due to the complexity of spectral data and the vastness of the chemical space. In this work, we introduce DiffNMR, a novel end-to-end framework that leverages a conditional discrete diffusion model for de novo molecular structure elucidation from NMR spectra. DiffNMR refines molecular graphs iteratively through a diffusion-based generative process, ensuring global consistency and mitigating error accumulation inherent in autoregressive methods.
The framework integrates a two-stage pretraining strategy that aligns spectral and molecular representations via diffusion autoencoder (Diff-AE) and contrastive learning, the incorporation of retrieval initialization and similarity filtering during inference, and a specialized NMR encoder with radial basis function (RBF) encoding for chemical shifts, preserving continuity and chemical correlation. Experimental results demonstrate that DiffNMR achieves competitive performance for NMR-based structure elucidation, offering an efficient and robust solution for automated molecular analysis.
\end{abstract}



\section{Introduction}
Rapid determination of molecular structures is expected to significantly accelerate the process of chemical discovery \cite{pollak2024development, liu2023assisting}. NMR spectroscopy is one of the most powerful techniques for determining molecular structures \cite{kuballa2023liquid, zhang2023detecting}. It provides insights into the local chemical environment of atoms within molecules, revealing details such as functional groups, bond connectivity, and stereochemistry through the magnetic resonance properties of atomic nuclei \cite{binev2007prediction, li2024highly}. As a result, NMR has become an indispensable analytical tool in fields such as synthetic chemistry and drug discovery. However, interpreting NMR spectra remains a notoriously challenging task. One major difficulty is spectral complexity, where a single peak may correspond to multiple chemical environments, further complicated by peak overlap and noise interference. Another challenge is the combinatorial explosion of possible structures. 
As the number of atoms increases, the number of chemically plausible structures grows exponentially, making exhaustive or manual searches impractical \cite{yang2023ultra}. NMR interpretation relies heavily on expert knowledge, which is time-consuming and prone to errors. While traditional Computer-Assisted Structure Elucidation tools \cite{elyashberg2021computer, elyashberg2023enhancing} can identify substructures or perform database matching \cite{lemm2024impact, sun2024cross}, their effectiveness is limited by the small size of reference spectral libraries (which only cover a fraction of known molecules) and the need for manual intervention. With the rapid advancement of automated synthesis and high-throughput experimentation, there is an urgent need to develop end-to-end methods capable of directly inferring molecular structures from NMR spectra without prior knowledge, thereby overcoming critical bottlenecks in the "synthesis-analysis" feedback loop.

In recent years, data-driven machine learning (ML) techniques have shown promise in NMR spectral analysis \cite{hu2023machine, kuhn2024nuclear, lu2024deep}, leveraging the increasing availability of spectral databases and simulated spectra. One class of ML approaches focuses on identifying molecular fragments from patterns in \hnmr\ and \cnmr\ NMR spectra and reconstructing complete structures through fragment assembly \cite{huang2021framework}. This approach emulates chemists' heuristic reasoning—piecing together a structure from known motifs. Nevertheless, the dependence on predefined fragment libraries and the combinatorial explosion of possible configurations with increasing molecular complexity hinder their applicability to larger systems. Reinforcement learning offers another angle \cite{kaelbling1996reinforcement,arulkumaran2017deep,sridharan2022deep, devata2023deepspinn}, where the spectral interpretation task is treated as a sequential decision-making process. A recent example is DeepSPInN \cite{devata2023deepspinn}, which formulates structure elucidation as a Markov decision process (MDP) and employs a Monte Carlo Tree Search (MCTS) guided by a neural network policy to iteratively build molecular graphs aligned with given spectra. For small molecules (<10 heavy atoms), it achieves high accuracy by leveraging multimodal spectral ( IR and \cnmr\ NMR) constraints. However, its practical utility remains limited by molecular complexity: MCTS exploration faces computational bottlenecks in larger systems due to the exponential scaling of decision paths, and performance degrades as structural intricacy increases.

Transformer-based \cite{han2021transformer, vaswani2017attention} deep learning models have recently been applied to NMR-based structure elucidation with notable success \cite{huang2021framework, hu2024accurate, alberts2024unraveling, alberts2023learning}. Transformers are sequence-to-sequence models adept at capturing long-range dependencies, which makes them well-suited to interpret the complex patterns in spectra and directly output a molecular structure. One such study presented a multitask transformer framework where the model simultaneously predicts substructure sets and complete molecular structure from raw \hnmr\ and \cnmr\ NMR spectra \cite{hu2024accurate}. By pretraining the transformer encoder-decoder to map 957 predefined substructures into SMILES strings, the model gains strong prior knowledge of chemical composition and achieved 69.6\% top-15 accuracy for molecules up to 19 heavy atoms. Another related transformer-based strategy capitalizing on the transformer’s strength in natural language tasks converts processed NMR spectral features into a textual sequence \cite{alberts2023learning}, where peak descriptors (chemical shift, multiplicity, intensity, etc.) extracted by NMR processing software (MestreNova) are formatted as tokens. A trained transformer encoder–decoder model translates this "NMR-speak" into a molecule's SMILES string and achieved 67\% top-1 accuracy with \hnmr\ and \cnmr\ NMR inputs on a million-scale dataset.

However, the autoregressive generation \cite{gregor2014deep} process of the transformer model has inherent limitations: the strict sequential sensitivity of generation can be "inconsistent with global interactions of atoms" in a molecule \cite{chen2024diffusion}, potentially leading to suboptimal structural decisions; moreover, even minor errors in the initial steps can cascade through the autoregressive chain, resulting in a drastically incorrect final structure or invalid SMILES. One promising direction to overcome these issues is the use of discrete graph diffusion models for molecule generation \cite{austin2021structured, vignac2022digress,wang2025madgen}. Diffusion models \cite{croitoru2023diffusion, ho2020denoising} represent a recently emerged class of generative models that construct the target (here, a molecular graph) through a gradual denoising process rather than a step-by-step assembly in a predetermined order. In a graph-based diffusion model, the entire molecule is refined collectively over many small steps, which means the generative process can inherently enforce global consistency at each refinement step.

Building on these insights, we propose a three-stage conditional diffusion framework, DiffNMR, to tackle the challenges in NMR-based structure elucidation. Our key contributions include: 
(1) a novel end-to-end framework, DiffNMR, utilizing a conditional discrete diffusion model for de novo molecular structure elucidation from NMR spectra.
(2) a specialized NMR encoder with radial basis function (RBF) encoding for chemical shifts, preserving continuity and enabling fine distinction between similar compounds.
(3) a two-stage pretraining strategy that first learns molecular representations via a diffusion autoencoder (diff-AE) and then aligns NMR spectral features with these representations using contrastive learning.
(4) a retrieval-initialized sampling scheme that enhances accuracy and efficiency by starting the diffusion process from database-retrieved structures similar to the input spectrum.


\begin{figure*}[!t]
    \centering
    \includegraphics[width=0.95\linewidth]{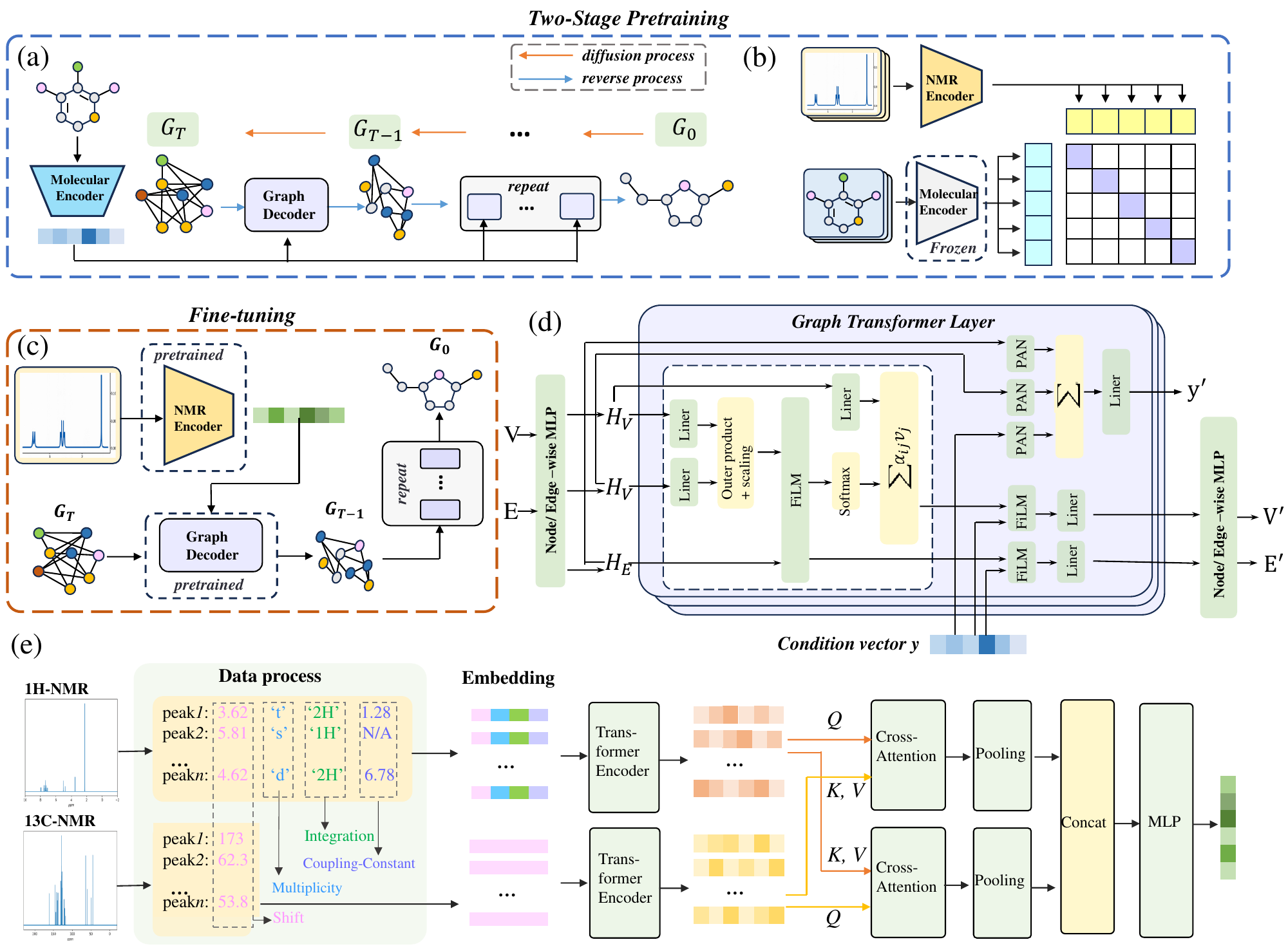} 
    \vspace{0pt}
    \caption{The framework of DiffNMR.
    \textbf{(a)} and \textbf{(b)} The two-stage pretraining process, in which the first pretraining stage \textbf{(a)} is a molecular diffusion autoencoder (Diff-AE) consisting of a molecular encoder and a graph decoder. The Diff-AE is based on the diffusion model, which defines two Markov processes: diffusion process and reverse process, and is designed to obtain a pretrained graph decoder and a pretrained molecular encoder capable of generating molecular representations. \textbf{(b)} The second pretraining stage—contrastive learning—is designed to align the NMR feature space with molecular representations space by training an NMR encoder  with molecular encoder keep frozen. \textbf{(c)} The fine-tuned end-to-end model, which integrating a pretrained NMR encoder and a pretrained diffusion decoder into a unified system for structure elucidation from NMR spectra. \textbf{(d)} The architecture of graph decoder. This denoising neural network is built on a graph transformer architecture and  employ Feature-wise Linear Modulation for condition generation. \textbf{(e)} The architecture of NMR encoder, which consist of domain-specific embedding layers, transformer-based modules and bidirectional cross-attention mechanism.}
    \label{fig:framework}
    \vspace{0pt}
\end{figure*}

\section{Result}


\subsection{Overview of DiffNMR}
The proposed DiffNMR is an end-to-end framework that leverages NMR spectral data as a strong conditional guide for de novo molecular structure generation. \cref{fig:framework} outlines an overview of the diffusion-based NMR elucidation framework, which illustrates its structural components and the processes involved in training and sampling.  

\textbf{The components of DiffNMR.} 
The framework comprises three key components: a molecular encoder, an NMR encoder, and a graph decoder. The molecular encoder, built on a graph transformer architecture\cite{min2022transformer, hu2020heterogeneous}, extracts feature vectors that encapsulate the structural and chemical properties of molecules. 
The graph decoder, adapted from the Digress model, is a denoising graph diffusion model that consists of a noise model and a denoising neural network. The noise model is a Markov process, involving successive graph edits (such as edge addition or deletion, and node or edge category modifications) that can occur independently on each node or edge. The denoising network $\phi_\theta(G^t, t)$ is trained to reverse this process by taking a noisy graph \( G^t = (V^t, E^t) \) as input and predicting the clean graph \( G^0 \). It is capable of controllable generation by taking the condition \( y \) as an additional input, i.e., $\phi_\theta(G^t, t, y)$. Specifically, this denoising network employs Feature-wise Linear Modulation (FiLM)\cite{brockschmidt2020gnn} in Graph Transformer layers, where condition-derived parameters adaptively affine-transform hidden features to align the generated graph distribution with \( y \)(see details in Method section).

Notably, we designed a specialized NMR encoder tailored to the characteristics of the provided dataset. Given the dataset’s structure—where \hnmr\ NMR peaks include chemical shift, splitting pattern, integral value, and coupling constants, while \cnmr\ NMR peaks consist solely of chemical shifts—the encoder processes these modalities through dedicated embedding and Transformer-based modules, followed by a fusion mechanism that integrates information from both modalities using bidirectional cross-attention. Domain-specific embedding layers transform raw spectral features into chemically meaningful vector representations: chemical shifts are encoded via radial basis functions (RBFs) to preserve continuous spatial relationships and local similarity, while discrete features (splitting patterns, integrals) and coupling constants are embedded using learnable matrices and dual-range RBF kernels, respectively. Separate Transformer encoders then model intra-modal peak relationships. A bidirectional cross-attention mechanism\cite{chen2021crossvit,wei2020multi} subsequently integrates \hnmr\ and \cnmr\ representations, enabling mutual enhancement by establishing complementary chemical context: coupling patterns refine carbon environment interpretation while carbon shifts guide proton signal assignment. This hierarchical design explicitly preserves spectroscopic topology compared to naive sequence-based encoding, prioritizing chemically critical interactions through structured feature embedding and multimodal attention(see details in Method section).

\textbf{Two-Stage Pretraining Strategy.} 
The training of DiffNMR unfolds in a two-stage pretraining strategy and is followed by a fine-tuning phase. 
In the first pretraining stage, a diffusion autoencoder \cite{preechakul2022diffusion,yang2023diffusion,fuest2024diffusion,abstreiter2021diffusion} is trained to establish a bidirectional mapping between molecular structures and their latent representations. The Diff-AE integrates a molecular encoder, which compresses molecular graphs into compact feature vectors, and a graph diffusion decoder, which reconstructs entity the original structures through an iterative denoising process. 
This diffusion process is inherently stochastic, leveraging randomness to explore a wide range of possible molecular configurations within the chemical space. 
By the end of this stage, we obtain a pretrained molecular encoder capable of producing compact, informative feature vectors and a pretrained conditional diffusion decoder adept at generating molecular structures.

The second pretraining stage aligns NMR spectra with molecular representations in the latent space. An NMR encoder is trained to extract feature vectors from NMR spectra that correspond to the molecular feature space defined in the first stage. This alignment is accomplished through contrastive learning \cite{you2020graph,le2020contrastive,yang2021cross}, where the NMR encoder minimizes the distance between feature vectors of matching spectrum-molecule pairs and maximizes it for non-matching pairs. The parameters of the molecular encoder remain fixed during this stage, compelling the NMR encoder to adapt to the established molecular representation space. This yields an NMR encoder that produces semantically meaningful spectral representations compatible with the diffusion decoder. Furthermore, this alignment facilitates efficient matching of spectra to molecular structures in databases, enabling both retrieval-initialized generation and similarity-based filtering during inference.

\textbf{Fine-tuning for Structure Elucidation.}
Following the pretraining stages, the pretrained NMR encoder and the pretrained diffusion decoder are unified into a cohesive end-to-end model and are jointly fine-tuned to establish a direct mapping from spectra to structures. The NMR encoder’s output serves as the conditioning signal for the diffusion decoder, guiding the denoising process toward generating molecules consistent with the input spectrum. This conditional generation leverages the decoder’s pretrained ability to sample chemically valid structures while incorporating spectral constraints, effectively balancing exploration (diverse sampling) and exploitation (structure fidelity).

\textbf{Retrieval Initialization and Similarity Filtering during Inference}
DiffNMR employs flexible sampling and filtering mechanisms to optimize performance, adapting seamlessly to scenarios with or without an external database.
When a database is accessible, DiffNMR uses a retrieval-initialized sampling scheme to enhance efficiency and accuracy. Leveraging the pretrained molecular and NMR encoders refined by contrastive learning, the framework retrieves a set of molecules from the database whose feature vectors closely align with the input NMR spectrum. These retrieved structures act as informed starting points for the diffusion process, reducing denoising steps and increasing the likelihood of convergence to the correct molecular structure. This method capitalizes on existing knowledge, proving particularly effective for complex spectra.
In the absence of a database, the generation process begins with a randomly initialized molecular structure, a hallmark of the diffusion model’s intrinsic randomness. This stochastic starting point allows the diffusion decoder to iteratively denoise and refine the structure, exploring a diverse array of possible candidate configurations before settling on a stable molecular form. While this flexibility broadens the exploration of the chemical space, it may require additional iterations to achieve precise results.
The randomness of the diffusion process generates multiple candidate structures, and contrastive learning provides a robust, quantitative method to select the optimal one. By computing cosine similarity scores between the feature vectors of generated candidates and the input NMR spectrum, DiffNMR ranks and filters these structures. This process discards less relevant candidates and prioritizes those most consistent with the spectrum, ensuring higher precision. The synergy of retrieval initialization and similarity-based filtering enhances DiffNMR’s adaptability and robustness across varying data scenarios.


\subsection{Dataset and Evaluation Metrics}

We conducted our experiments on the Multimodal Spectroscopic Dataset (MSD), which comprises 794,403 unique molecules derived from the USPTO reaction dataset \cite{alberts2024unraveling}. These molecules include chemical species containing carbon, hydrogen, oxygen, nitrogen, sulfur, phosphorus, silicon, boron, and halogens. Each molecule contains between 5 and 35 heavy atoms and is associated with five simulated complementary spectra: \hnmr NMR, \cnmr NMR, HSQC‑NMR, Infrared, and Mass spectra. NMR simulations were performed using MestReNova with deuterated chloroform as the solvent and default acquisition parameters. Specifically, the dataset provides detailed NMR annotations, including chemical shifts, multiplet analysis (e.g., singlet, doublet, triplet), normalized peak integrations and coupling constants for \hnmr NMR. Likewise, chemical shifts and relative intensities are provided for proton-decoupled \cnmr NMR spectra, offering comprehensive spectral information essential for molecular structure elucidation. 

We evaluate the performance of our models using two metrics: Top-$k$ accuracy and Tanimoto similarity. Top-$k$ accuracy quantifies the probability that the correct target structure appears among the Top-$k$ predicted molecular candidates, while Tanimoto similarity measures the structural similarity between the predicted molecules and the ground truth based on molecular fingerprints. A higher Tanimoto score indicates a closer match to the true structure. To compute this metric, we used the RDKit toolkit to generate Morgan fingerprints (Morgan, 1965) \cite{morgan1965generation} with a radius of 2 and a bit length of 2048. 

\subsection{DiffNMR enables effective structure elucidation}
\begin{table}[h]
    \setlength{\abovecaptionskip}{7pt}
    \renewcommand{\arraystretch}{1.5}
    \setlength{\tabcolsep}{4pt} 
    \centering
    \begin{tabular}{c c c c c c c c c c c c c}
        \hline \hline
        & & \multicolumn{3}{c}{$\le15$} & \multicolumn{3}{c}{$\le20$} & \multicolumn{3}{c}{$\le25$} \\
        \cmidrule(lr){3-5} \cmidrule(lr){6-8} \cmidrule(lr){9-11}
        & Formula  & Top-1 & Top-5 & Top-10 & Top-1 & Top-5 & Top-10 & Top-1 & Top-5 & Top-10 \\
        \hline
        \multirow{3}[-10]{*}{\hnmr\ NMR} & $\surd$ & 58.56\% & 65.54\% & 69.79\% & 57.48\% & 65.09\% & 69.85\% & 52.04\% & 61.04\% & 65.75\% \\
        \multirow{3}[-10]{*}{\cnmr\ NMR} & $\surd$ & 27.75\% & 36.34\% & 42.46\% & 22.73\% & 32.92\% & 40.04\% & 19.75\% & 29.94\% & 37.33\% \\
        \multirow{3}[-10]{*}{\hnmr\ +\cnmr\ NMR} & $\surd$ & \textbf{68.26}\% & \textbf{75.04}\% & \textbf{80.27}\% & \textbf{67.10}\% & \textbf{74.98}\% & \textbf{79.58}\% & \textbf{61.55}\% & \textbf{70.05}\% & \textbf{75.07}\% \\
        \hline
        \multirow{3}[-10]{*}{\hnmr NMR} & $\times$ & 39.17\% & 46.69\% & 51.73\% & 45.68\% & 52.53\% & 56.28\% & 40.91\% & 48.40\% & 52.34\% \\
        \multirow{3}[-10]{*}{\cnmr NMR} & $\times$ & 14.61\% & 22.02\% & 26.12\% & 14.56\% & 21.58\% & 26.62\% & 11.60\% & 18.50\% & 22.69\% \\
        \multirow{3}[-10]{*}{\hnmr\ +\cnmr\ NMR} & $\times$ & \textbf{58.83}\% & \textbf{66.61}\% & \textbf{71.39}\% & \textbf{60.34}\% & \textbf{67.30}\% & \textbf{71.40}\% & \textbf{53.10}\% & \textbf{61.36}\% & \textbf{65.53}\% \\
        \hline \hline
    \end{tabular}
    \caption{Prediction accuracies (Top-1, Top-5, Top-10) for molecular structures based on different NMR inputs (\hnmr NMR, \cnmr NMR, \hnmr\ +\cnmr\ NMR) in datasets with $\le15$,  $\le20$, and $\le25$ heavy atoms.}
    \label{table-1}
\end{table}

Our experiments, summarized in \Cref{table-1}, evaluate model performance on molecular structure prediction across datasets with $\le15$,  $\le20$, and $\le25$ heavy atoms, comparing the effects of molecular formula inclusion and different NMR input types (\hnmr, \cnmr, and \hnmr\ +\cnmr).
It's obvious that the prediction accuracy is enhanced when the molecular formula is provided compared to cases where it is absent. For datasets with up to 15 heavy atoms, the Top-1 accuracy for the combined \hnmr\ + \cnmr\ NMR increases from 58.83\% (without formula) to 68.26\% (with formula). This improvement remains consistent across datasets with $\le20$ and $\le25$ heavy atom. The boost in accuracy arises from the structural constraint imposed by the molecular formula, which  specifies elemental composition and possible isomers, thus effectively reducing combinatorial complexity. 

Furthermore, the comparison of input modalities reveals distinct advantages of \hnmr\ NMR over \cnmr\ NMR, as well as the synergistic benefits of combining both spectra. For example, under the formula-included condition ($\le15$ atoms), the Top-1 accuracy of \hnmr\ NMR (58.56\%) substantially surpasses that of \cnmr\ NMR (27.75\%), primarily due to the richer structural details encoded in \hnmr\ NMR, such as coupling patterns, integration values, and splitting multiplicity compared to the fewer, yet backbone-specific, carbon signals, which directly reflect hydrogen connectivity and local environments. Additionally, the highest performance is achieved when \hnmr\ + \cnmr\ NMR spectra are jointly utilized, with a Top-1 accuracy of 68.26\%. This synergy arises because \cnmr\ NMR complements \hnmr\ NMR by resolving ambiguities in carbon skeleton and distinguishing functional groups with overlapping proton shifts.  They together forming a more complete spectral profile that resolves structural ambiguities effectively. The superior accuracy of the combined \hnmr\ + \cnmr\ NMR input underscores that our NMR encoder effectively encodes and fuses information from both spectra, significantly enhancing the model’s predictive power. Given this advantage, our subsequent experiments will focus exclusively on the \hnmr\ + \cnmr\ NMR input configuration.

Moreover, the accuracy across datasets with heavy atom counts of $\le15$, $\le20$, and $\le25$ highlights a intricate balance between molecular complexity and dataset size. For the combined \hnmr\ + \cnmr\ NMR(without formula), Top-1 accuracy changes from 58.83\% ($\le15$ atoms) to 60.34\% ($\le20$ atoms) and further to 53.10\% ($\le25$ atoms). As the heavy atom count rises from 15 to 20, training data expands from approximately 120,000 to 270,000 structures (a $\sim$1.4-fold increase), and model performance remains largely stable, suggesting that the added data sufficiently accommodates the increased structural complexity. However, when the count increases from 20 to 25, the training data grows from about 270,000 to 390,000 structures (a $\sim$0.5-fold increase), yet accuracy drops by roughly 7\%. This decrease indicates that the modest data increase fails to keep pace with the exponential growth in structural diversity at higher atom counts, resulting in lower data density per structural variant. This trend reveals that a larger and more diverse dataset would further elevate the model’s performance, especially for larger molecules. It is worth noting that the results shown in Table 1 only utilized similarity filtering in the reasoning stage and did not employ the retrieval initialization strategy. Once the retrieval initialization strategy is adopted, the decline in accuracy caused by the increase in the number of atoms will be alleviated, changing from a $\sim$7\% decrease to a $\sim$2\% decrease (further information can be found in section 2.4).

\begin{figure*}[!t]
    \centering
    \includegraphics[width=1\linewidth]{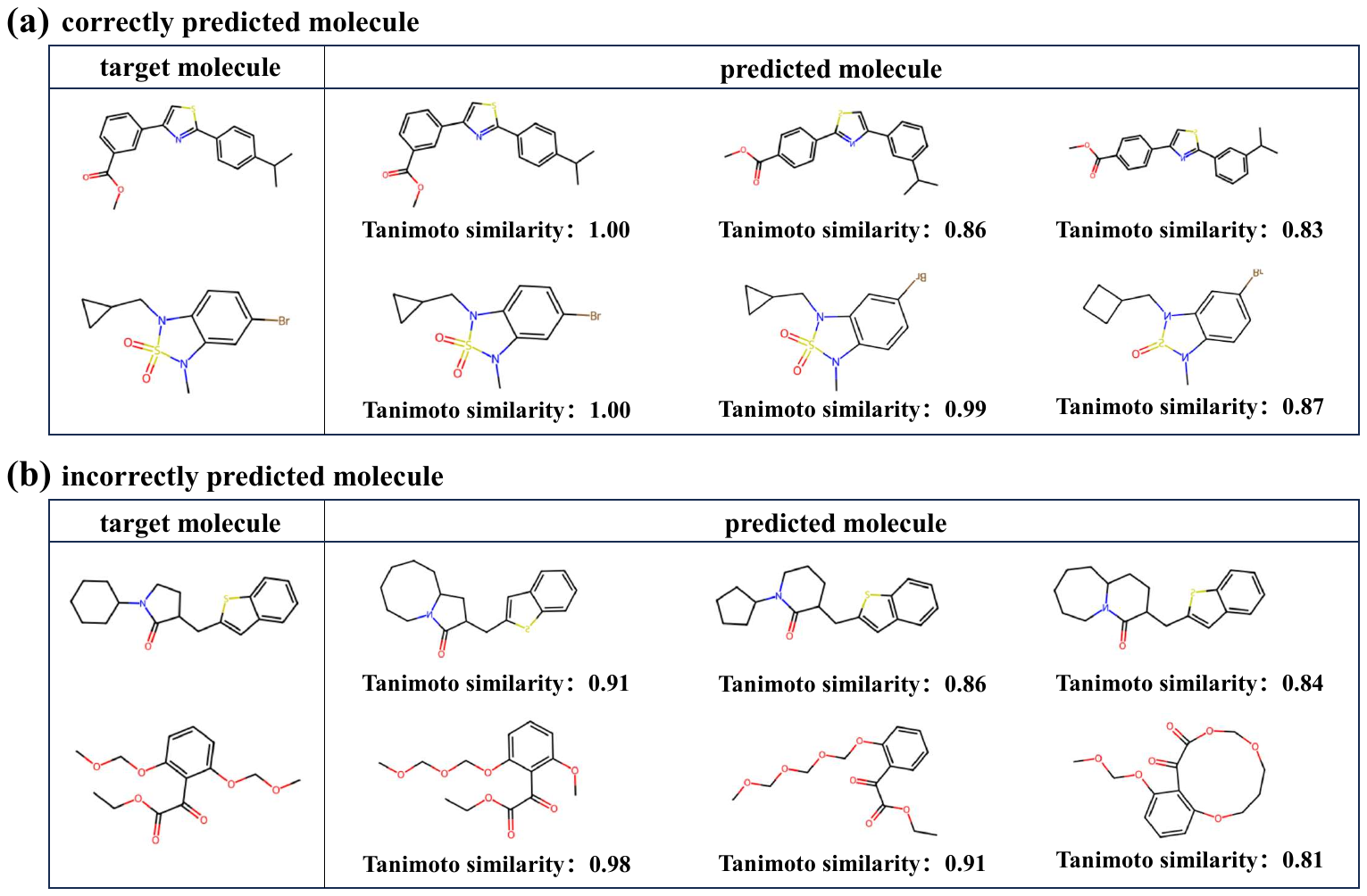} 
    \vspace{-50pt}
    \caption{Examples of top-3 sampled molecular structures with their Tanimoto similarity scores to the target molecule. (a) Correctly predicted molecules: The first predicted molecule in each set matches the target perfectly (Tanimoto similarity = 1), while the other two predictions have slightly lower similarities. (b) Incorrectly predicted molecules: None of the top-3 predicted molecules match the target exactly, with Tanimoto similarities less than 1}
    \label{fig:example}
    \vspace{0pt}
\end{figure*}

\begin{figure*}[!t]
    \centering
    \includegraphics[width=0.8\linewidth]{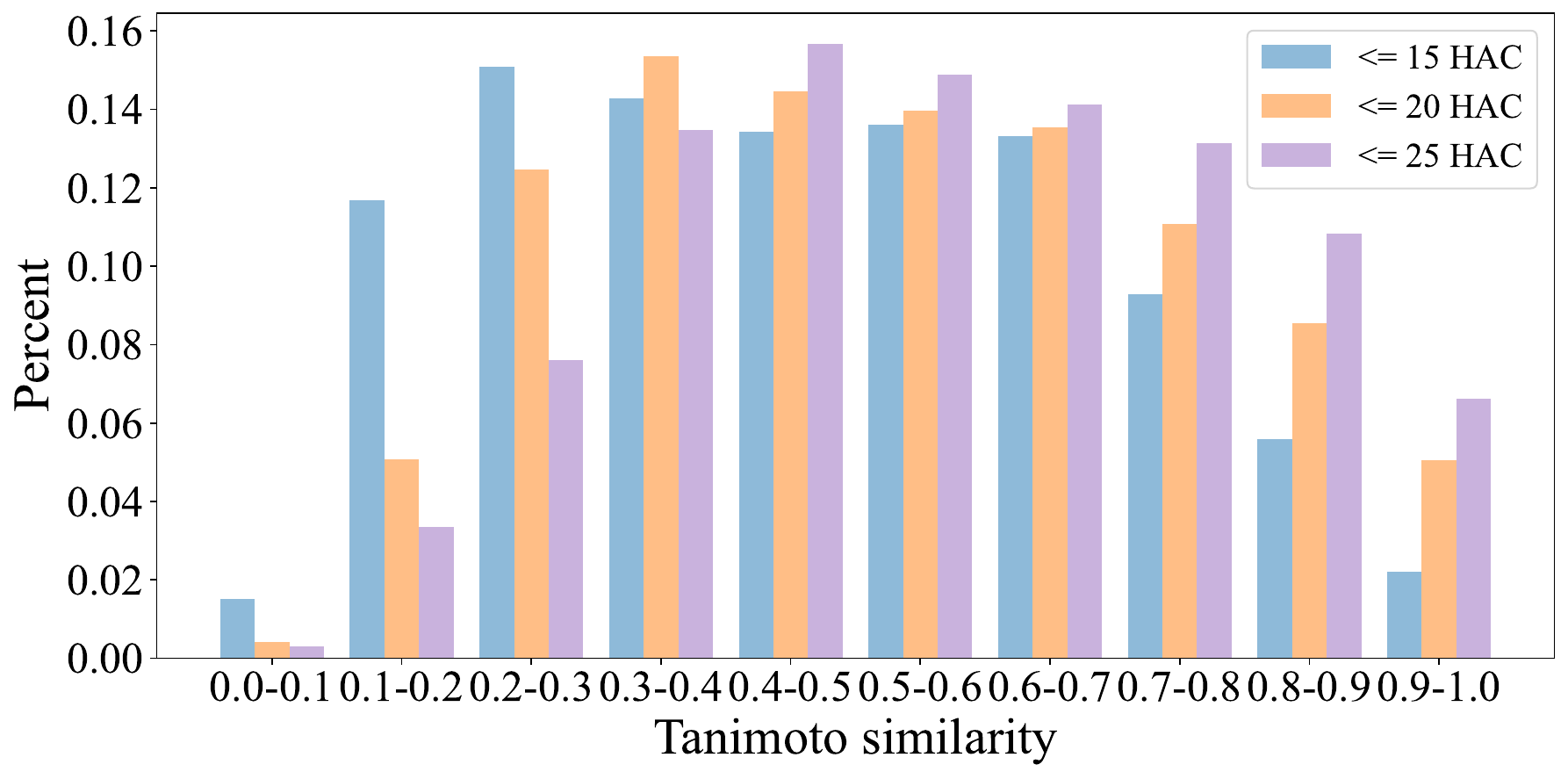} 
    \vspace{0pt}
    \caption{Distribution of Tanimoto similarity scores for incorrect Top-1 predictions (excluding exact matches) across molecules with heavy atom counts $\le15$, $\le20$, and $\le25$. The invalid SMILES strings and the correctly-predicted molecules were excluded.}
    \label{fig:distribution}
    \vspace{0pt}
\end{figure*}

\cref{fig:example}(a) provides two illustrative examples of correctly predicted molecules. In both cases, the top-1 ranked prediction matches the target molecule exactly (Tanimoto similarity = 1.00), demonstrating the model’s ability to resolve structural connectivity with high fidelity. Notably, In cases where the model fails to rank the correct molecule within the top-3 predictions, as shown in \cref{fig:example}(b), the the predicted structures demonstrate high Tanimoto similarity scores (0.81–0.98) relative to the target molecules. These scores reflect a significant degree of structural overlap, as the predicted molecules preserve key chemical motifs, such as functional groups and bonding patterns, anticipated in the target structures. This retention of critical features suggests that, even in failure, the model generates outputs that are structurally relevant and can serve as useful starting points for further refinement or manual analysis of molecular connectivity. Such behavior underscores the model’s robustness, offering valuable insights into the system under study despite not achieving an exact match.

The distribution of Tanimoto similarities for incorrect Top-1 predictions, stratified by heavy atom count (HAC), is presented in \cref{fig:distribution}. Strikingly, the fraction of predictions with Tanimoto similarity $\ge0.5$ increases with molecular complexity: 44.0\% ($\le15$ HAC), 52.2\% ($\le20$ HAC), and 59.6\% ($\le25$ HAC), and the similarity distribution shifts toward higher values as HAC increases. This trend can be attributed to two factors. First, larger molecules often exhibit more distinct spectral fingerprints due to their extended carbon frameworks and diverse functional groups, enabling the model to capture broader structural patterns. Second, the training data for high-HAC molecules ($\sim$390k for $\le25$ HAC) likely encompass recurring motifs (e.g., polycyclic systems or branched chains), which the model learns to prioritize during generation. However, the combinatorial explosion of isomerism at higher HACs introduces challenges in resolving fine-grained structural details, leading to minor deviations in substituent placement or stereochemistry. The observed trends in \cref{fig:example}(b) and \cref{fig:distribution} have practical implications for the model’s application. While exact structure prediction remains the ultimate goal, the ability to produce high-similarity outputs in failure cases enhances the model’s value as a tool for chemists.

\subsection{Similarity Filtering and Retrieval Initialization Enhance Structure Elucidation}

\begin{table}[h]
    \setlength{\abovecaptionskip}{7pt}
    \renewcommand{\arraystretch}{1.5}
    \setlength{\tabcolsep}{4pt} 
    \centering
    \begin{tabular}{c c c c c c c c c c}
        \hline \hline
        & & \multicolumn{2}{c}{$\le15$} & \multicolumn{2}{c}{$\le20$} & \multicolumn{2}{c}{$\le25$} \\
        \cmidrule(lr){3-4} \cmidrule(lr){5-6} \cmidrule(lr){7-8}
        \makecell[c]{Similarity-based \\ filtering} & \makecell[c]{Retrieval \\ initialization} & Accuracy & \makecell[c]{Tanimoto \\ Similarity}  & Accuracy & \makecell[c]{Tanimoto \\ Similarity} & Accuracy & \makecell[c]{Tanimoto \\ Similarity} \\
        \hline
        \multirow{3}[-10]{*}{$\times$} & $\times$ & 51.11\% & 0.73 & 43.25\% & 0.72 & 35.76\% & 0.71 \\
        \multirow{3}[-10]{*}{$\times$} & $\surd$ & 53.87\% & 0.75 & 53.07\% & 0.78 & 49.70\% & 0.78 \\
        \hline
        \multirow{3}[-10]{*}{$\surd$} & $\times$ & 58.83\% & 0.78 & 60.34\% & 0.81 & 53.10\% & 0.79 \\
        \multirow{3}[-10]{*}{$\surd$} & $\surd$ & \textbf{60.13}\% & \textbf{0.79} & \textbf{61.18}\% & \textbf{0.82} & \textbf{58.47}\% & \textbf{0.82} \\
        \hline \hline
    \end{tabular}
    \caption{Performance metrics for DiffNMR without molecular formulas, showing Top-1 accuracy and average Tanimoto similarity under varying conditions of similarity-based filtering and retrieval initialization.}
    \label{table-2}
\end{table}

\begin{figure*}[!t]
    \centering
    \includegraphics[width=0.6\linewidth]{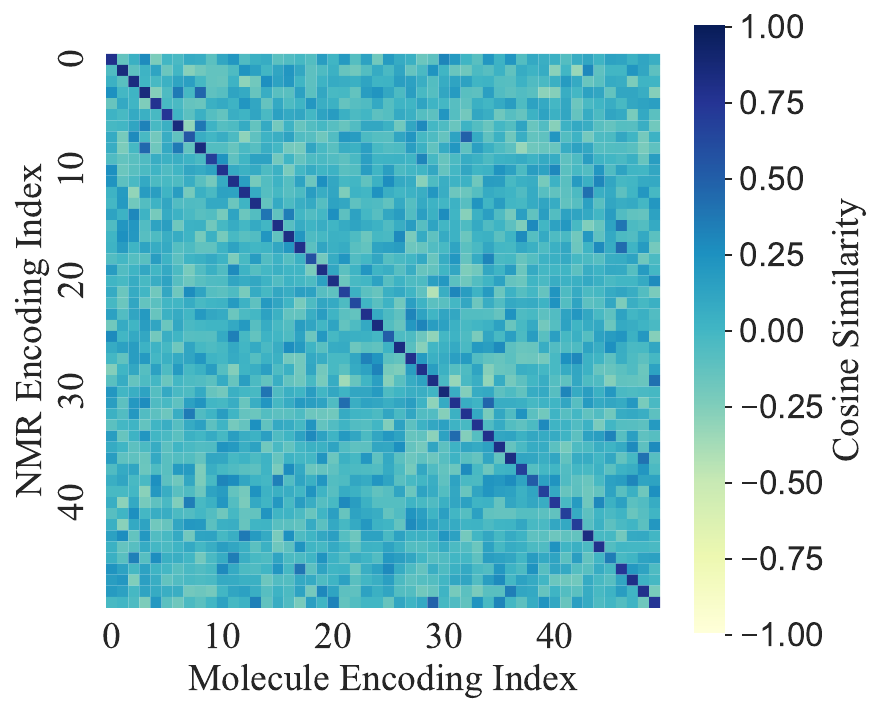} 
    \vspace{-5pt}
    \caption{Heatmap showing cosine similarity scores for 50 randomly selected NMR-molecule pairs from the test set. The diagonal corresponds to matching NMR-molecule pairs, and off-diagonal elements represent non-matching pairs.}
    \label{fig:consin similarity}
    \vspace{0pt}
\end{figure*}

\begin{figure*}[!t]
    \centering
    \includegraphics[width=1\linewidth]{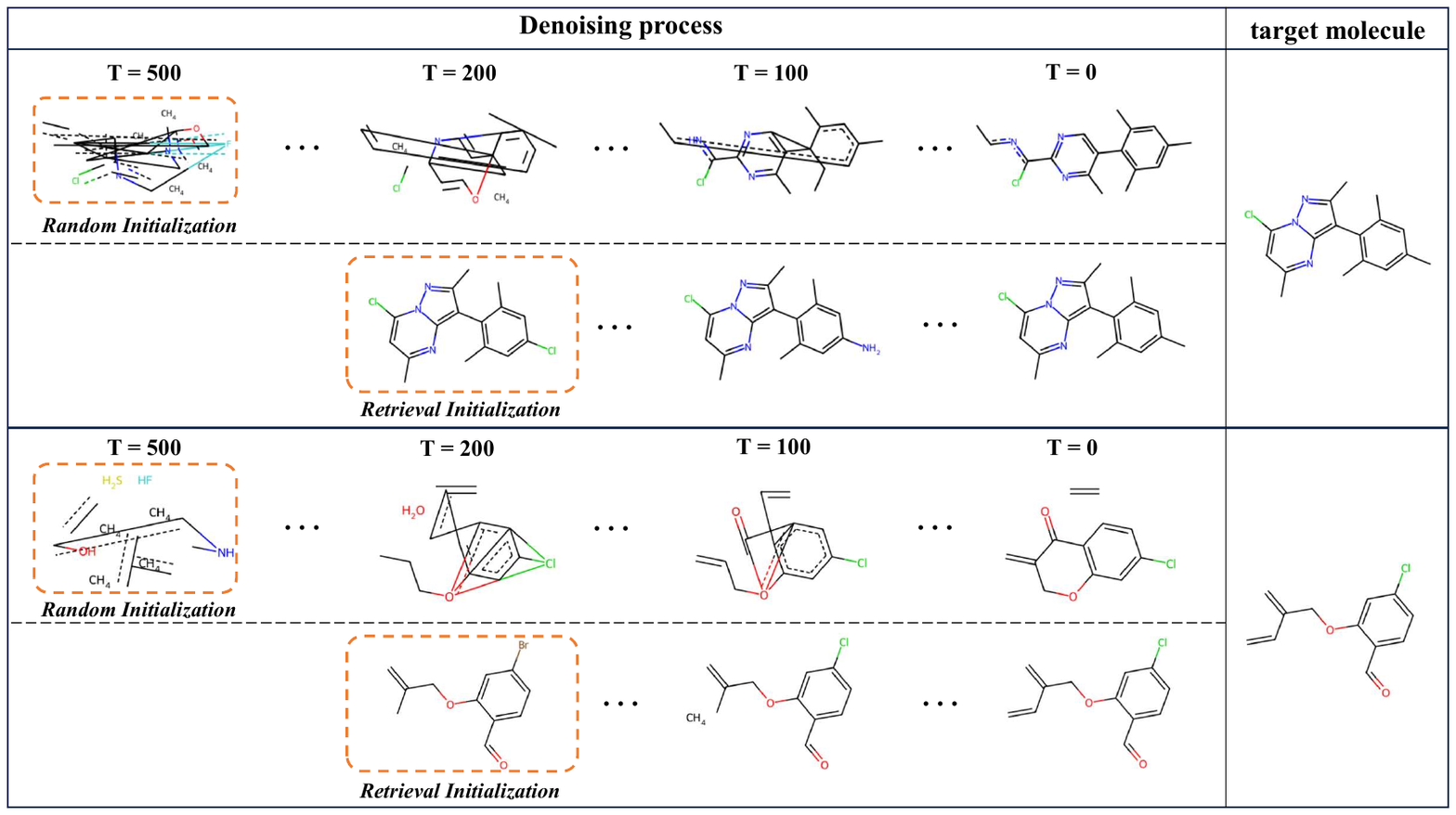} 
    \vspace{-50pt}
    \caption{Examples of structure prediction for two target molecules using two inference strategies. (i) Random initialization starts from a random graph and performs 500 denoising steps, resulting in incorrect predictions. (ii) Retrieval Initialization selects a similar molecule from the database based on cosine similarity between NMR-encoded and molecular representations, followed by 200 denoising steps, yielding correct predictions matching the target molecules.}
    \label{fig:retrival example}
    \vspace{0pt}
\end{figure*}

To further enhance the performance of DiffNMR in molecular structure elucidation, we evaluate two key strategies—similarity-based filtering and retrieval initialization—employed in inference, which leveraging pretrained representations from contrastive learning to refine the structure generation process. Table 2 presents the Top-1 accuracy and average Tanimoto similarity across datasets with different heavy atom counts. Here, we only consider the scenario without molecular formulas.

DiffNMR computes cosine similarity scores between feature vectors of generated molecular candidates and the input NMR spectrum, ranking and filtering candidates to pinpoint the structure most consistent with the spectral data. This approach yields notable improvements, as seen in \Cref{table-2}. For the $\le15$ HAC dataset, Top-1 accuracy rises from 51.11\% (without filtering) to 58.83\% (with filtering) when no additional initialization is applied. For larger molecules in the $\le25$ HAC dataset, accuracy jumps from 35.76\% to 53.10\%, and the average Tanimoto similarity also increases, from 0.71 to 0.79. These consistent gains across datasets underscore the ability of similarity-based filtering to effectively select the most spectrally aligned structure from a diverse pool of candidates, thereby boosting prediction reliability. 

This similarity-based filtering capability is underpinned by the contrastive learning pretraining, as demonstrated in \cref{fig:consin similarity}. We randomly selected 50 NMR-molecule pairs from the test set and computed cosine similarity scores between their encoded representations, visualized as a heatmap. The diagonal entries—representing matching NMR-molecule pairs—consistently show the highest similarity scores, approaching 1.0, while off-diagonal entries (non-matching pairs) exhibit significantly lower values. This pronounced separation confirms that contrastive learning successfully aligns NMR spectra with their corresponding molecular structures in the latent space, providing a robust foundation for similarity-based filtering. By leveraging this alignment, DiffNMR can accurately rank and filter candidates, ensuring that the top prediction closely matches the input spectrum, particularly in scenarios with multiple plausible structures. 

Additionally, the diffusion process benefits from retrieval initialization, which leverages a database to provide informed starting points for the diffusion process, markedly improving prediction accuracy, especially for larger molecules. Here, we use the training set as the retrieval database. According to \Cref{table-2}, when similarity-based filtering is enabled, retrieval initialization increases the Top-1 accuracy from 58.83\% to 60.13\% in the $\le15$ HAC dataset. This enhancement becomes more substantial as molecular complexity rises: for the $\le25$ HAC dataset, accuracy improves from 53.10\% to 58.47\%. Notably, the accuracy gap between smaller ($\le15$ HAC) and larger ($\le25$ HAC) molecules—previously observed as a ~7\% decline in section 2.2—narrows to a ~2\% decrease with retrieval initialization. This mitigation of the performance drop suggest that retrieval initialization constrains the diffusion process with chemically relevant initial structures, effectively bridging the accuracy disparity caused by increased molecular complexity.

\cref{fig:retrival example} illustrates how Retrieval Initialization leverages prior knowledge from the database to narrow the search space through two examples of structure elucidation for target molecules. For each target, two inference approaches are compared: (i) starting from a randomly initialized graph, which, after 500 denoising steps, yields an incorrect prediction, and (ii) using retrieval initialization, where a molecule with a feature vector most similar to the input NMR spectrum (determined by cosine similarity) is retrieved from the database as the starting point, leading to the correct structure after only 200 denoising steps. This initializing with a spectroscopically informed structure enables the model to refine a near-correct starting point rather than exploring blindly. Furthermore,  these examples reveal that retrieval initialization not only enhances the accuracy but also accelerates sampling efficiency by reducing the required denoising steps. This efficiency gain is critical for larger molecules, where random initialization struggles most due to the vast combinatorial space of possible structures, this informed approach is particularly impactful.

The combined application of similarity-based filtering and retrieval initialization maximizes DiffNMR’s performance.  In the $\le25$ HAC dataset, Top-1 accuracy(without formulas) reaches 58.47\% compared to 35.76\% without either strategy, and average Tanimoto similarity climbs from 0.71 to 0.82. This improvement of performance stems from informed starting points narrowing the search space and similarity-based filtering ensuring precise final selections. Together, these strategies enhance both the precision and structural relevance of the generated molecules.

\subsection{Evaluate the effectiveness of pretraining and NMR encoder}

To assess the effectiveness of the proposed two-stage pretraining strategy in the DiffNMR framework, we conducted experiments on the dataset with $\leq15$ heavy atoms, evaluating the impact of pretraining on Top-1 prediction accuracy. The first-stage pretraining Diff-AE provides pretrained parameters for the graph decoder, while the second-stage pretraining contrastive learning provides pretrained parameters for the NMR encoder. 
We compared the Top-1 accuracy in four scenarios: no pretraining applied to either component, pretraining applied solely to the graph decoder using Diff-AE, pretraining applied only to the NMR encoder via contrastive learning, and pretraining applied to both components together. 
These experiments were conducted without retrieval initialization or similarity filtering during the inference stage. The results, depicted in Figure 6, show that pretraining either component individually improves the accuracy by approximately 5\% over the no-pretraining baseline, while pretraining both components yields a synergistic improvement of about 10\%. These findings confirm the effectiveness of the proposed two-stage pretraining approach in enhancing the DiffNMR framework's performance.

\begin{figure*}[!h]
    \centering
    \includegraphics[width=0.9\linewidth]{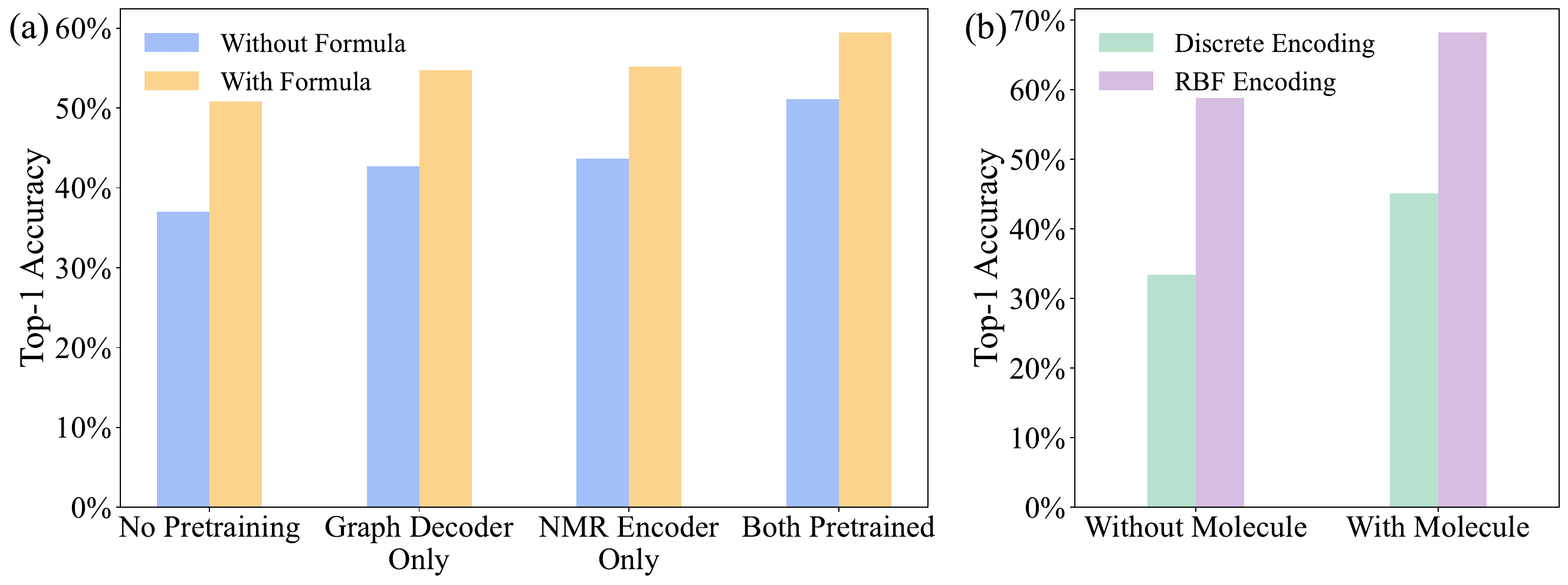} 
    \vspace{-5pt}
    \caption{Top-1 accuracy of molecular structure prediction for molecules with up to 15 heavy atoms, with and without molecular formula inclusion. (a) Comparison of four pretraining scenarios: no pretraining, pretraining only the graph decoder, pretraining only the NMR encoder, and pretraining both components. (b) Comparison of RBF encoding and discrete encoding adopted in NMR encoder.}
    \label{fig:ablation_pretraing}
    \vspace{0pt}
\end{figure*}

Next, we shift our attention to the NMR encoder’s encoding scheme and its role in enhancing molecular structure elucidation to evaluate the effectiveness of the NMR encoder within the DiffNMR framework. Chemical shifts in NMR spectral data are continuous variables that reflect the electronic environment of atoms and are critical for structural prediction. We analyze the impact of two encoding methods for these shifts: discrete encoding, which treats them as categorical values, and RBF encoding, which leverages radial basis functions to represent them as continuous values (see Section 4.3 for details). As shown in Figure 6(a), RBF encoding significantly improves Top-1 prediction accuracy by over 20\% compared with discrete encoding in scenarios both with and without molecular formulas, demonstrating its superior ability to capture structural information from NMR spectra. The effectiveness of RBF encoding is likely to stem from its ability to embed chemical shifts in a continuous vector space where proximity reflects similarity.  By using radial basis functions, RBF encoding maps shifts such that those with close numerical values are positioned nearer to each other in the embedding space. This preservation of continuity ensures that the NMR encoder can detect and leverage small differences in shifts, which are often pivotal for distinguishing between structurally similar compounds. For instance, consider two molecules with nearly identical carbon skeletons but differing by a single functional group. Their chemical shifts might differ by only a very litter, a nuance that RBF encoding captures by placing these shifts in close proximity within the vector space. This allows the model to recognize their structural similarity while retaining the resolution needed to differentiate them. However, discrete encoding treats chemical shifts as isolated categories, discarding their continuity and relational information. By assigning each shift to a separate vector, discrete encoding fails to capture the proximity between similar shifts, resulting in a loss of sensitivity to these fine structural distinctions.


\section{Conclusion and Discussion}
In this work, we presented DiffNMR, a novel framework for molecular structure elucidation from NMR spectra, leveraging a conditional discrete diffusion model. The framework features a specialized NMR encoder with RBF encoding, which captures the continuity of chemical shift and proximity between similar shifts, which is critical for identifing and distinguishing similar compounds, and a two-stage pretraining strategy that aligns molecular and spectral representations via a diffusion autoencoder and contrastive learning. The incorporation of retrieval initialization and similarity filtering further boosts performance, particularly for complex molecules, reducing the accuracy gap as molecular size increases. Unlike autoregressive models, which suffer from sequential sensitivity and error propagation, DiffNMR employs an iterative diffusion process that refines molecular graphs holistically, ensuring global structural consistency. Our experimental results validate DiffNMR’s effectiveness, achieving competitive performance relative to autoregressive approaches.

Despite its strengths, DiffNMR’s performance diminishes with larger molecules due to limited data density, emphasizing the need for broader, more diverse training datasets to address greater structural variability. Future research could investigate advanced diffusion techniques or hybrid models to enhance scalability and accuracy, integrate additional spectral modalities such as 2D NMR to better resolve structural ambiguities, or extend the framework to address stereochemical features. In conclusion, DiffNMR offers a promising framework for automating NMR-based structure elucidation, with the potential to revolutionize spectral analysis by delivering faster, more precise compound identification. Furthermore, integrating automated NMR processing into high-throughput workflows could significantly improve efficiency and accelerate discoveries in chemical research.

\section{Method}
\label{sec:others}

\subsection{Notation and problem formulation}
We represent a molecule as a graph \( G = (V, E) \), where \( V \) denotes atoms with attributes in a discrete space \( \mathcal{V} \), and \( E \) denotes bonds with attributes in \( \mathcal{E} \). Node attributes are encoded in a matrix \( \mathbf{V} \in \mathbb{R}^{N \times a} \), where \( N \) is the number of atoms, \( a = \dim(\mathcal{V}) \), and each row \( \mathbf{v}_i \) is a one-hot vector for atom \( i \). Edge attributes are encoded in a sparse tensor \( \mathbf{E} \in \mathbb{R}^{N \times N \times b} \), with \( b = \dim(\mathcal{E}) \), where \( \mathbf{e}_{ij} \) indicates the bond type between atoms \( i \) and \( j \). We denote $G^{t}$ for $t = 1, \ldots, T$ as a sequence of noisy graphs where $t$ indicates the index of diffusion steps. 
NMR interpretation can be formalized as molecular generation under strict spectral constraints, where the goal is to model the true conditional distribution $q(G|y)$ where $\mathbf{y}$ denotes conditional signal. To achieve this, we parameterize the distribution as $p_{\theta}(G|y)$ and implement it using a conditional denoising network $\phi_{\theta}(G_t,t,y)$, which leverages $y$ to guide the denoising process of a noisy graph aligning the molecule with the NMR.

\subsection{Discrete diffusion for conditional graph generation}
Diffusion models consist of forward diffusion and reverse denoising processes. The diffusion process $ q(G^{1:T} | G^0) = \prod_{t=1}^T q(G^t | G^{t-1})$ corrupts molecular graph $G=G^0=(V^0, E^0)$ into noisy versions $G^t=(V^t, E^t)$, and as $T \to \infty$, $q(G^T)$ converges to a stationary distribution $\pi(G)$. The reverse process $p_\theta(G^{0:T}) = q(G^T) \prod_{t=1}^T p_\theta(G^{t-1} | G^t)$, parameterized by $\theta$, reconstructs the graph from the noisy states $G^t$.
The diffusion process perturbs \( G^0 \) in a discrete state-space to capture molecular structure. Transition matrices $\mathbf{Q}_V \in \mathbb{R}^{a \times a}$ and $\mathbf{Q}_E \in \mathbb{R}^{b \times b}$ govern the noise for the nodes $V$ and edges $E$ of the graph, respectively. Each diffusion step is defined as $q(G^t | G^{t-1}) = (\mathbf{V}^{t-1} \mathbf{Q}_V^t, \mathbf{E}^{t-1} \mathbf{Q}_E^t)$. As the process is Markovian, the noisy states $G^t$ can be built from $G^0$ using

\begin{equation}
    q(G^t | G^0) = (\mathbf{V} \bar{\mathbf{Q}}_V^t, \mathbf{E} \bar{\mathbf{Q}}_E^t),
\end{equation}

for $\bar{\mathbf{Q}}_V^t = \mathbf{Q}_V^1 \dots \mathbf{Q}_V^t$ and $\bar{\mathbf{Q}}_E^t = \mathbf{Q}_E^1 \dots \mathbf{Q}_E^t$. Vignac et al. defined $\pi(G) = (\mathbf{m}_X \in \mathbb{R}^{a}, \mathbf{m}_E \in \mathbb{R}^{b})$ as the marginal distributions of atom types and bond types, and

\begin{equation}
    \bar{\mathbf{Q}}^t = \alpha^t \mathbf{I} + (1 - \alpha^t) \mathbf{1} \mathbf{m}',
\end{equation}

for atoms or bonds, where $\mathbf{m}'$ denotes the transposed row vector. The cosine schedule is typically chosen for $\alpha^t = \cos\left(0.5 \pi \left( \frac{t}{T} + s \right) / (1 + s)\right)^2$.

The reverse process, which generates  $G^0$ iteratively conditioned on $y$ starting with an initial sample $G^T \sim \pi(G^t)$, can be implemented by estimating the distribution on the graph  $p_\theta(G^{t-1 }| G^t, y)$ using denoising neural network $\phi_{\theta}(G_t,t,y)$. This distribution can be modeled as a product over nodes and edges:
\begin{equation}
\begin{aligned}
    p_\theta(G^{t-1} \mid G^t,y) = \prod_{1 \leq i \leq n} p_\theta(v_i^{t-1} \mid G^t,y) \prod_{1 \leq i, j \leq n} p_\theta(e_{ij}^{t-1} \mid G^t,y)
\end{aligned}
\end{equation}
where  $p_\theta(v_i^{t-1} \mid G^t,y)$ and  $ p_\theta(e_{ij}^{t-1} \mid G^t,y)$ are respectively marginalized over predictions of node and edge types:

\begin{equation}
\begin{aligned}
   p_\theta(v_i^{t-1} \mid G^t,y) &=  \sum_{v \in \mathcal{V}} p_\theta(v_i^{t-1} \mid v_i = v, G^t,y) p_{\theta}(v_i = v \mid G^t,y)\\
   p_\theta(e_{ij}^{t-1} \mid G^t,y) &=  \sum_{e \in \mathcal{E}} p_\theta(e_{ij}^{t-1} \mid e_{ij} = v, G^t,y) p_{\theta}(e_{ij} = e \mid G^t,y)
\end{aligned}
\end{equation}
in which $p_\theta(v_i^{t-1} \mid v_i = v, G^t,y)=q(v_i^{t-1} \mid v_i = v)$ and $p_\theta(e_{ij}^{t-1} \mid e_{ij} = e, G^t,y)=q(e_{ij}^{t-1} \mid e_{ij} = e)$. 

The denoising neural network $\phi_{\theta}(G^t,t,y)$ with trainable parameters $\theta$ takes both noisy graph $G^t=(\mathbf{V}^t, \mathbf{E}^t)$ and condition $y$ as inputs, and aims to predict the clean graph $G^0$. This neural network $\phi_{\theta}$ could be trained to minimize the cross-entropy loss $\mathcal{L}$ between the predicted probabilities $p_{\theta}(G^0 \mid G^t,y)$ and the true graph $G^0$:

\begin{equation}
\begin{aligned}
    \mathcal{L}(G^0,p_{\theta}(G^0 \mid G^t,y)) & = \sum_{1 \leq i \leq n} \text{CE}(\mathbf{v}_i,p_{\theta}(v_i \mid G^t,y)) \\
    & + \lambda \sum_{1 \leq i,j \leq n} \text{CE}(\mathbf{e}_{ij}, p_{\theta}(e_{ij} \mid G^t,y)),
\end{aligned}
\end{equation}
where $\lambda$ determines the weighting between the contributions of nodes and edges . 

\subsection{Contrastive Learning}

In the second pretraining stage, a contrastive learning strategy is utilized to train an NMR encoder, aligning NMR spectra with molecular representations, so that the generated NMR spectral feature vectors closely correspond to the molecular feature space established in the first stage. This is accomplished by designating matching spectrum-molecule pairs as positive and non-matching pairs as negative. The NMR encoder is optimized to reduce the difference between feature vectors of positive pairs while increasing the difference for negative pairs. During this process, the molecular encoder's parameters remain unchanged, allowing the NMR encoder to adapt to the existing molecular representation framework.

The training is driven by a contrastive loss function based on Noise Contrastive Estimation (InfoNCE), defined as:

\begin{equation}
\mathcal{L}_{\text{contrastive}} = -\frac{1}{N} \sum_{i=1}^{N} \log \frac{\exp(\text{score}(\mathbf{z}_i, \mathbf{m}_i) / \tau)}{\sum_{j=1}^{N} \exp(\text{score}(\mathbf{z}_i, \mathbf{m}_j) / \tau)}
\end{equation}

Here, $N$ represents the batch size, $\mathbf{z}_i$ is the feature vector from the $i$-th NMR spectrum extracted by the NMR encoder, $\mathbf{m}_i $ is the corresponding molecular feature vector from the fixed molecular encoder, $\text{score}(\cdot, \cdot)$ denotes the similarity metric, and $\tau$ is the temperature hyperparameter.

The similarity between vectors is computed using a dot-product-based score:

\begin{equation}
\text{score}(\mathbf{z}_i, \mathbf{m}_j) = \frac{\mathbf{z}_i^\top \mathbf{m}_j}{\sqrt{|\mathbf{z}_i|^2 |\mathbf{m}_j|^2}}
\end{equation}

This score prioritizes directional alignment, enabling the NMR encoder to capture semantic relationships between spectra and molecules effectively. Through this contrastive learning approach, the NMR encoder generates spectral representations that are both meaningful and compatible with molecular features, facilitating integration with the diffusion decoder and enabling efficient spectrum-to-molecule retrieval during inference.

\subsection{NMR Encoder}
In this work, we present an NMR spectral encoder designed to capture the intricate chemical information embedded in NMR spectra, This encoder architecture comprises two key components: (1) Feature embedding layers with domain-specific encoding schemes, and (2) Bidirectional cross-attention layer that fuses information from both \hnmr\ and \cnmr\ NMR.

The embedding layer processes the diverse features of NMR spectra, including chemical shifts, splitting patterns, integral values, and coupling constants for \hnmr\ NMR, and chemical shifts for \cnmr\ NMR. For chemical shifts $\delta$, which represent the resonance frequency of nuclei relative to a reference and reflect the local electronic environment, we employ a Radial Basis Function (RBF) encoding. This encoding is defined as:

\begin{equation}
\text{RBF}(\delta) = \exp \left( -\frac{(\delta - c_i)^2}{2\sigma^2} \right), \quad i = 1, \ldots, B
\end{equation}

where $c_i$ are centers linearly spaced across the chemical shift range, $B$ is the number of bins, and $\sigma$ is an adaptive bandwidth ensuring resolution across the range. This RBF encoding maps each $\delta$ value to a vector by computing its similarity to a set of predefined centers using Gaussian kernels, where shifts proximate to a center $c_i$ yield higher values in the corresponding dimension. This embedding schemes preserve the continuity chemical shift and their sensitivity to subtle structural variations which are pivotal for distinguishing between similar compounds.
Unlike discrete embeddings, the RBF encoding maintains proximity in the embedding space for chemically similar shifts, enabling it to generalize across structurally related molecules while retaining the resolution necessary to capture these fine distinctions. Similarly, the coupling constants $J$ in \hnmr\ NMR, which quantify the strength of spin-spin interactions, are also continuous and appear as variable-length sets per peak. We apply RBF encoding to each coupling constant individually, using two ranges with distinct resolution to cover typical values, then sum the resulting vectors to produce a fixed-size representation. 

Additionally, for other NMR features, such as splitting patterns $S$ (e.g., "s" for singlet, "d" for doublet, etc.), and integral values $I$ (e.g., "1H", "2H", etc.) in \hnmr\ NMR, we utilize standard 
learnable embedding layers:

\begin{equation}
\mathbf{e}_{\text{split}} = \text{Emb}(\text{idx}(S)), \quad \mathbf{e}_{\text{integral}} = \text{Emb}(\text{idx}(I))
\end{equation}

where $\text{idx}(S)$ and $\text{idx}(I)$ respectively maps splitting descriptors and integral values to predefined indices. These features are categorical or integer-based, wherein splitting patterns arise from spin-spin coupling, reflecting the number and type of neighboring nuclei, and integral values indicating the number of equivalent nuclei contributing to a peak in \hnmr\ NMR. Assigning dense, learnable vectors to these features allows the model to learn distinct representations for each category while capturing latent relationships during training, leveraging the chemical significance of each category without imposing a continuous structure.

The embeddings for all features of a peak in \hnmr\ NMR are concatenated, and then the final \hnmr and \cnmr NMR embedding are:

\begin{equation}
\mathbf{e}_{H} = \text{Concat}( \text{RBF}(\delta_h), \mathbf{e}_{\text{split}}, \mathbf{e}_{\text{integral}}, \text{RBF}(J) ), \quad \mathbf{e}_{C}  = \text{RBF}(\delta_c)
\end{equation}

Following these embeddings,  the vector $\mathbf{e}_{H}$ and $\mathbf{e}_{C}$ are separately encoded through standard Transformer encoder layers: 

\begin{equation}
\mathbf{h}_H = \text{Transformer}(\mathbf{e}_H), \mathbf{h}_C = \text{Transformer}(\mathbf{e}_C)
\end{equation}

These Transformer encoders model intra-spectral relationships, such as coupling networks in \hnmr\ spectra and carbon skeleton connectivity in \cnmr\ spectra, via self-attention mechanisms.

\hnmr NMR provides insights into hydrogen environments and their coupling, while \cnmr NMR reveals the carbon skeleton of the molecule. To integrate the complementary information from \hnmr\ and \cnmr\ NMR spectra, we incorporate a bidirectional cross-attention mechanism:

\begin{equation}
\begin{aligned}
\text{attn}_{H \rightarrow C} &= \text{softmax} \left( \frac{\mathbf{Q}_H \mathbf{K}_C^\top}{\sqrt{d_k}} \right) \mathbf{V}_C \\
\text{attn}_{C \rightarrow H} &= \text{softmax} \left( \frac{\mathbf{Q}_C \mathbf{K}_H^\top}{\sqrt{d_k}} \right) \mathbf{V}_H
\end{aligned}
\end{equation}

Here,  $\mathbf{Q}_{H/C}$, $\mathbf{K}_{C/H}$, $\mathbf{V}_{C/H}$ are the query, key, and value matrices derived from the representations of \hnmr\ and \cnmr\ spectra , respectively. The outputs, $\text{attn}_{H \rightarrow C}$
and $\text{attn}_{C \rightarrow H}$ represent the \hnmr peaks informed by \cnmr peaks and vice versa.
This bidirectional cross-attention ensures mutual enhancement, allowing the model to learn correlations between these spectra—such as the correspondence between a proton’s chemical shift and the carbon it is attached to, which are chemically significant and can reveal a more holistic view of the molecular structure. 
The attention outputs are concatenated with the original hidden representations, and then pooled to produce modality-specific representations:

\begin{equation}
\mathbf{g}_{H} = \text{AvgPool}( \text{Concat}[\mathbf{h}_H, \text{attn}_{H \rightarrow C})]), \quad \mathbf{g}_{C}  = \text{AvgPool}( \text{Concat}[\mathbf{h}_C, \text{attn}_{C \rightarrow H}])
\end{equation}

Finally, these \hnmr\ and \cnmr\ spectra representations are concatenated and projected through MLP to produce the ultimate NMR spectra encoding:

\begin{equation}
\mathbf{F}_{\text{NMR}} = \text{MLP}( \text{Concat}[\mathbf{g}_H, \text{g}_{C})]).
\end{equation}

This NMR encoder integrates RBF embeddings for NMR features with a Transformer-based architecture and bidirectional cross-attention, producing a rich, robust representation of NMR spectra that enhances molecular structure analysis.

\subsection{Graph Decoder}
We employ a graph transformer network[] as denoising network to take a noisy graph $G^t=(V^t, E^t)$ as input and predict a distribution over corresponding clean graphs. The network begins by encoding node, edge, global features and conditional vector using an MLP-based encoder. These encoded features are then processed through a sequence of graph transformer layers (GTLs).  Each GTL integrates graph structure and NMR condition signal. Within each GTL, the node features are first refined using self-attention and then combined with the edge features via the FiLM operation, defined as:

\begin{equation}
\text{FiLM}(\mathbf{X}_1, \mathbf{X}_2) = \mathbf{X}_1 \mathbf{W}_1 + (\mathbf{X}_1 \mathbf{W}_2) \odot \mathbf{X}_2 + \mathbf{W}_2
\end{equation}

where $W_1$ and $W_2$ are learnable parameters. The edge features are updated based on attention scores global features and conditional vector. The global features and conditional vector are updated by fusing themselves with aggregated node and edge features obtained through the PNA operation, which is given by:

\begin{equation}
\text{PNA}(\mathbf{X}) = \text{Concat}(\text{max}(\mathbf{X}), \text{min}(\mathbf{X}), \text{mean}(\mathbf{X}), \text{std}(\mathbf{X})) \mathbf{W}
\end{equation}

where $W$ is a learnable parameter. To ensure training stability, each GTL incorporates layer normalization and fully-connected layers. Finally, the refined node and edge features are passed through an MLP-based decoder to produce the final graph representation. 

\section{Code Availability}
This model is developed on the PaddlePaddle which is parallel distribution deep learning framework, and PaddleMaterial which is data-driven and mechanism-driven toolkit for materials R\&D.
The code of DiffNMR is available at \url{https://github.com/PaddlePaddle/PaddleMaterial}

\section{Acknowledgements}
This work is supported by the National Science and Technology Major Project (Grants No.2023ZD0120702) and Basic Research Program of Jiangsu (BK20231215)

\bibliographystyle{unsrt}  
\bibliography{references}  

\begin{thebibliography}{10}

\bibitem{pollak2024development}
Julie Pollak, Moses Mayonu, Lin Jiang, and Bo~Wang.
\newblock The development of machine learning approaches in two-dimensional nmr data interpretation for metabolomics applications.
\newblock {\em Analytical Biochemistry}, 695:115654, 2024.

\bibitem{liu2023assisting}
Sirui Liu, Haotian Chu, Yuhao Xie, Fangming Wu, Ningxi Ni, Chenghao Wang, Fangjing Mu, Jiachen Wei, Jun Zhang, Mengyun Chen, et~al.
\newblock Assisting and accelerating nmr assignment with restrained structure prediction.
\newblock {\em bioRxiv}, pages 2023--04, 2023.

\bibitem{kuballa2023liquid}
Thomas Kuballa, Katja~H Kaltenbach, Jan Teipel, and Dirk~W Lachenmeier.
\newblock Liquid nuclear magnetic resonance (nmr) spectroscopy in transition—from structure elucidation to multi-analysis method.
\newblock {\em Separations}, 10(11):572, 2023.

\bibitem{zhang2023detecting}
Congcong Zhang, Li~Xu, Qingxia Huang, Yulan Wang, and Huiru Tang.
\newblock Detecting submicromolar analytes in mixtures with a 5 min acquisition on 600 mhz nmr spectrometers.
\newblock {\em Journal of the American Chemical Society}, 145(47):25513--25517, 2023.

\bibitem{binev2007prediction}
Yuri Binev, Maria~MB Marques, and Jo{\~a}o Aires-de Sousa.
\newblock Prediction of 1h nmr coupling constants with associative neural networks trained for chemical shifts.
\newblock {\em Journal of chemical information and modeling}, 47(6):2089--2097, 2007.

\bibitem{li2024highly}
Jie Li, Jiashu Liang, Zhe Wang, Aleksandra~L Ptaszek, Xiao Liu, Brad Ganoe, Martin Head-Gordon, and Teresa Head-Gordon.
\newblock Highly accurate prediction of nmr chemical shifts from low-level quantum mechanics calculations using machine learning.
\newblock {\em Journal of chemical theory and computation}, 20(5):2152--2166, 2024.

\bibitem{yang2023ultra}
Qiong Yang, Hongchao Ji, Zhenbo Xu, Yiming Li, Pingshan Wang, Jinyu Sun, Xiaqiong Fan, Hailiang Zhang, Hongmei Lu, and Zhimin Zhang.
\newblock Ultra-fast and accurate electron ionization mass spectrum matching for compound identification with million-scale in-silico library.
\newblock {\em Nature Communications}, 14(1):3722, 2023.

\bibitem{elyashberg2021computer}
Mikhail Elyashberg and Dimitris Argyropoulos.
\newblock Computer assisted structure elucidation (case): current and future perspectives.
\newblock {\em Magnetic Resonance in Chemistry}, 59(7):669--690, 2021.

\bibitem{elyashberg2023enhancing}
Mikhail Elyashberg, Sriram Tyagarajan, Mihir Mandal, and Alexei~V Buevich.
\newblock Enhancing efficiency of natural product structure revision: leveraging case and dft over total synthesis.
\newblock {\em Molecules}, 28(9):3796, 2023.

\bibitem{lemm2024impact}
Dominik Lemm, Guido~Falk von Rudorff, and O~Anatole Von~Lilienfeld.
\newblock Impact of noise on inverse design: the case of nmr spectra matching.
\newblock {\em Digital Discovery}, 3(1):136--144, 2024.

\bibitem{sun2024cross}
Hanyu Sun, Xi~Xue, Xue Liu, Hai-Yu Hu, Yafeng Deng, and Xiaojian Wang.
\newblock Cross-modal retrieval between 13c nmr spectra and structures based on focused libraries.
\newblock {\em Analytical Chemistry}, 96(15):5763--5770, 2024.

\bibitem{hu2023machine}
Guilin Hu and Minghua Qiu.
\newblock Machine learning-assisted structure annotation of natural products based on ms and nmr data.
\newblock {\em Natural Product Reports}, 40(11):1735--1753, 2023.

\bibitem{kuhn2024nuclear}
Stefan Kuhn, R{\^o}mulo~Pereira de~Jesus, and Ricardo~Moreira Borges.
\newblock Nuclear magnetic resonance and artificial intelligence.
\newblock {\em Encyclopedia}, 4(4):1568--1580, 2024.

\bibitem{lu2024deep}
Xin-Yu Lu, Hao-Ping Wu, Hao Ma, Hui Li, Jia Li, Yan-Ti Liu, Zheng-Yan Pan, Yi~Xie, Lei Wang, Bin Ren, et~al.
\newblock Deep learning-assisted spectrum--structure correlation: state-of-the-art and perspectives.
\newblock {\em Analytical Chemistry}, 96(20):7959--7975, 2024.

\bibitem{huang2021framework}
Zhaorui Huang, Michael~S Chen, Cristian~P Woroch, Thomas~E Markland, and Matthew~W Kanan.
\newblock A framework for automated structure elucidation from routine nmr spectra.
\newblock {\em Chemical Science}, 12(46):15329--15338, 2021.

\bibitem{kaelbling1996reinforcement}
Leslie~Pack Kaelbling, Michael~L Littman, and Andrew~W Moore.
\newblock Reinforcement learning: A survey.
\newblock {\em Journal of artificial intelligence research}, 4:237--285, 1996.

\bibitem{arulkumaran2017deep}
Kai Arulkumaran, Marc~Peter Deisenroth, Miles Brundage, and Anil~Anthony Bharath.
\newblock Deep reinforcement learning: A brief survey.
\newblock {\em IEEE Signal Processing Magazine}, 34(6):26--38, 2017.

\bibitem{sridharan2022deep}
Bhuvanesh Sridharan, Sarvesh Mehta, Yashaswi Pathak, and U~Deva Priyakumar.
\newblock Deep reinforcement learning for molecular inverse problem of nuclear magnetic resonance spectra to molecular structure.
\newblock {\em The Journal of Physical Chemistry Letters}, 13(22):4924--4933, 2022.

\bibitem{devata2023deepspinn}
Sriram Devata, Bhuvanesh Sridharan, Sarvesh Mehta, Yashaswi Pathak, Siddhartha Laghuvarapu, Girish Varma, and Deva Priyakumar.
\newblock Deepspinn-multimodal deep learning for molecular structure prediction from infrared and nmr spectra.
\newblock 2023.

\bibitem{han2021transformer}
Kai Han, An~Xiao, Enhua Wu, Jianyuan Guo, Chunjing Xu, and Yunhe Wang.
\newblock Transformer in transformer.
\newblock {\em Advances in neural information processing systems}, 34:15908--15919, 2021.

\bibitem{vaswani2017attention}
Ashish Vaswani, Noam Shazeer, Niki Parmar, Jakob Uszkoreit, Llion Jones, Aidan~N Gomez, {\L}ukasz Kaiser, and Illia Polosukhin.
\newblock Attention is all you need.
\newblock {\em Advances in neural information processing systems}, 30, 2017.

\bibitem{hu2024accurate}
Frank Hu, Michael~S Chen, Grant~M Rotskoff, Matthew~W Kanan, and Thomas~E Markland.
\newblock Accurate and efficient structure elucidation from routine one-dimensional nmr spectra using multitask machine learning.
\newblock {\em ACS Central Science}, 10(11):2162--2170, 2024.

\bibitem{alberts2024unraveling}
Marvin Alberts, Oliver Schilter, Federico Zipoli, Nina Hartrampf, and Teodoro Laino.
\newblock Unraveling molecular structure: A multimodal spectroscopic dataset for chemistry.
\newblock {\em Advances in Neural Information Processing Systems}, 37:125780--125808, 2024.

\bibitem{alberts2023learning}
Marvin Alberts, Federico Zipoli, and Alain~C Vaucher.
\newblock Learning the language of nmr: Structure elucidation from nmr spectra using transformer models.
\newblock 2023.

\bibitem{gregor2014deep}
Karol Gregor, Ivo Danihelka, Andriy Mnih, Charles Blundell, and Daan Wierstra.
\newblock Deep autoregressive networks.
\newblock In {\em International Conference on Machine Learning}, pages 1242--1250. PMLR, 2014.

\bibitem{chen2024diffusion}
Hongyang Chen, Can Xu, Lingyu Zheng, Qiang Zhang, and Xuemin Lin.
\newblock Diffusion-based graph generative methods.
\newblock {\em IEEE Transactions on Knowledge and Data Engineering}, 2024.

\bibitem{austin2021structured}
Jacob Austin, Daniel~D Johnson, Jonathan Ho, Daniel Tarlow, and Rianne Van Den~Berg.
\newblock Structured denoising diffusion models in discrete state-spaces.
\newblock {\em Advances in neural information processing systems}, 34:17981--17993, 2021.

\bibitem{vignac2022digress}
Clement Vignac, Igor Krawczuk, Antoine Siraudin, Bohan Wang, Volkan Cevher, and Pascal Frossard.
\newblock Digress: Discrete denoising diffusion for graph generation.
\newblock {\em arXiv preprint arXiv:2209.14734}, 2022.

\bibitem{wang2025madgen}
Yinkai Wang, Xiaohui Chen, Liping Liu, and Soha Hassoun.
\newblock Madgen--mass-spec attends to de novo molecular generation.
\newblock {\em arXiv preprint arXiv:2501.01950}, 2025.

\bibitem{croitoru2023diffusion}
Florinel-Alin Croitoru, Vlad Hondru, Radu~Tudor Ionescu, and Mubarak Shah.
\newblock Diffusion models in vision: A survey.
\newblock {\em IEEE Transactions on Pattern Analysis and Machine Intelligence}, 45(9):10850--10869, 2023.

\bibitem{ho2020denoising}
Jonathan Ho, Ajay Jain, and Pieter Abbeel.
\newblock Denoising diffusion probabilistic models.
\newblock {\em Advances in neural information processing systems}, 33:6840--6851, 2020.

\bibitem{min2022transformer}
Erxue Min, Runfa Chen, Yatao Bian, Tingyang Xu, Kangfei Zhao, Wenbing Huang, Peilin Zhao, Junzhou Huang, Sophia Ananiadou, and Yu~Rong.
\newblock Transformer for graphs: An overview from architecture perspective.
\newblock {\em arXiv preprint arXiv:2202.08455}, 2022.

\bibitem{hu2020heterogeneous}
Ziniu Hu, Yuxiao Dong, Kuansan Wang, and Yizhou Sun.
\newblock Heterogeneous graph transformer.
\newblock In {\em Proceedings of the web conference 2020}, pages 2704--2710, 2020.

\bibitem{brockschmidt2020gnn}
Marc Brockschmidt.
\newblock Gnn-film: Graph neural networks with feature-wise linear modulation.
\newblock In {\em International Conference on Machine Learning}, pages 1144--1152. PMLR, 2020.

\bibitem{chen2021crossvit}
Chun-Fu~Richard Chen, Quanfu Fan, and Rameswar Panda.
\newblock Crossvit: Cross-attention multi-scale vision transformer for image classification.
\newblock In {\em Proceedings of the IEEE/CVF international conference on computer vision}, pages 357--366, 2021.

\bibitem{wei2020multi}
Xi~Wei, Tianzhu Zhang, Yan Li, Yongdong Zhang, and Feng Wu.
\newblock Multi-modality cross attention network for image and sentence matching.
\newblock In {\em Proceedings of the IEEE/CVF conference on computer vision and pattern recognition}, pages 10941--10950, 2020.

\bibitem{preechakul2022diffusion}
Konpat Preechakul, Nattanat Chatthee, Suttisak Wizadwongsa, and Supasorn Suwajanakorn.
\newblock Diffusion autoencoders: Toward a meaningful and decodable representation.
\newblock In {\em Proceedings of the IEEE/CVF conference on computer vision and pattern recognition}, pages 10619--10629, 2022.

\bibitem{yang2023diffusion}
Xingyi Yang and Xinchao Wang.
\newblock Diffusion model as representation learner.
\newblock In {\em Proceedings of the IEEE/CVF International Conference on Computer Vision}, pages 18938--18949, 2023.

\bibitem{fuest2024diffusion}
Michael Fuest, Pingchuan Ma, Ming Gui, Johannes Schusterbauer, Vincent~Tao Hu, and Bjorn Ommer.
\newblock Diffusion models and representation learning: A survey.
\newblock {\em arXiv preprint arXiv:2407.00783}, 2024.

\bibitem{abstreiter2021diffusion}
Korbinian Abstreiter, Sarthak Mittal, Stefan Bauer, Bernhard Sch{\"o}lkopf, and Arash Mehrjou.
\newblock Diffusion-based representation learning.
\newblock {\em arXiv preprint arXiv:2105.14257}, 2021.

\bibitem{you2020graph}
Yuning You, Tianlong Chen, Yongduo Sui, Ting Chen, Zhangyang Wang, and Yang Shen.
\newblock Graph contrastive learning with augmentations.
\newblock {\em Advances in neural information processing systems}, 33:5812--5823, 2020.

\bibitem{le2020contrastive}
Phuc~H Le-Khac, Graham Healy, and Alan~F Smeaton.
\newblock Contrastive representation learning: A framework and review.
\newblock {\em Ieee Access}, 8:193907--193934, 2020.

\bibitem{yang2021cross}
Zhuo Yang, Jianfei Song, Minjian Yang, Lin Yao, Jiahua Zhang, Hui Shi, Xiangyang Ji, Yafeng Deng, and Xiaojian Wang.
\newblock Cross-modal retrieval between 13c nmr spectra and structures for compound identification using deep contrastive learning.
\newblock {\em Analytical Chemistry}, 93(50):16947--16955, 2021.

\bibitem{morgan1965generation}
Harry~L Morgan.
\newblock The generation of a unique machine description for chemical structures-a technique developed at chemical abstracts service.
\newblock {\em Journal of chemical documentation}, 5(2):107--113, 1965.

\end{thebibliography}






\end{document}